\documentclass[preprint,aps]{revtex4}
\usepackage{color}
\usepackage{mathtools}
\usepackage{diagbox}
\usepackage{appendix}
\usepackage{amsmath,amssymb,amsfonts,dcolumn,color,graphicx,graphics,latexsym,epsfig}
\usepackage[compatibility=false]{caption}
\usepackage{subcaption}
\usepackage{hyperref}
\usepackage{natbib}
\usepackage{float}
\usepackage{booktabs}
\def\beq{\begin{equation}}
\def\eeq{\end{equation}}
\def\barr{\begin{array}}
\def\earr{\end{array}}

\usepackage{caption}
\usepackage{subcaption}

\begin{document}
\title{Analog Raychaudhuri equation in mechanics}

\author{Rajendra Prasad Bhatt\footnote{{Present Address: Inter-University Centre for Astronomy and Astrophysics, Post Bag 4, Ganeshkhind, Pune 411 007, India}}, Anushree Roy and
Sayan Kar}
\email{bhattrajendra1997@gmail.com, anushree@phy.iitkgp.ac.in, sayan@phy.iitkgp.ac.in}
\affiliation{Department of Physics, Indian Institute of Technology
Kharagpur, 721 302, India}

\begin{abstract}
\noindent Usually, in mechanics, we obtain the trajectory of a
particle in a given force field by solving Newton's second law with
chosen initial conditions. In contrast, through our work here,
we first demonstrate how one may analyse the behaviour of a 
suitably defined family
of trajectories of a given mechanical system. 
Such an approach leads us to develop a mechanics analog following the
well-known Raychaudhuri equation largely studied in Riemannian  geometry and general 
relativity. The idea of geodesic focusing,  
which is more familiar to a relativist, 
appears to be analogous to
the meeting of trajectories of a mechanical system within a finite time. 
Applying our general results to the case of simple pendula,
we obtain relevant quantitative consequences. Thereafter, we set up 
and perform a straightforward experiment based on a system 
with two pendula.
The experimental results on this system are found to tally well with
our proposed theoretical model. In summary, the simple theory, as
well as the related  experiment, provides us with
a way to understand the essence of a fairly involved 
concept in advanced physics 
from an elementary standpoint.

\end{abstract}

\maketitle

\section{\bf Introduction}
\begin{figure}[h!]
\epsfxsize=4in\epsffile{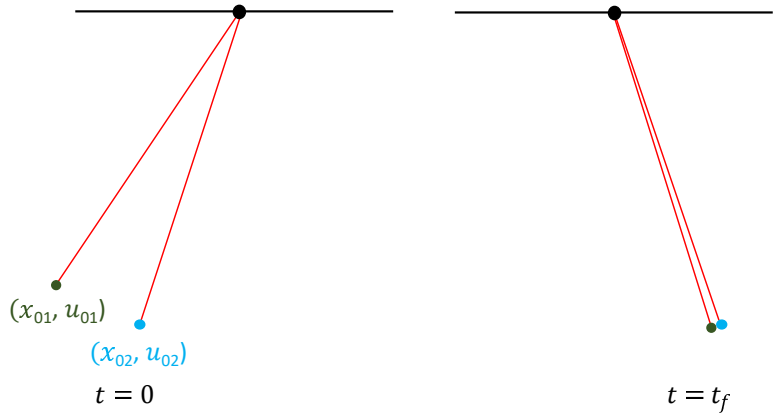} \caption{{Simple pendula: Here $x_{01}, x_{02}$ are initial positions and $u_{01}, u_{02}$ are  initial velocities of the two bobs (figure on the left). Meeting of the bobs at $t=t_f$ (figure on the right).}} \label{diagram1}
\end{figure}

\noindent Imagine two pendula of the same length hung from a common support.
Let us give different initial displacements to the bobs and set them in motion 
(in a single vertical plane) with
different initial velocities (see Figure \ref{diagram1}). It is obvious that they will strike each other
after a finite time. What does this time of striking/meeting depend upon?
How does one develop a general theoretical model for such scenarios (with several
simple pendula or for other systems) and also
set up a simple experiment? Our aim, in this article, revolves around such
issues and questions which, to the best of our knowledge, have not been addressed in standard texts on mechanics \cite{symon,kk}. In particular, we concentrate on 
the collective behaviour of
families of trajectories of a given mechanical system. 

\noindent  It turns out that such studies are directly related to 
the well-known Raychaudhuri equation \cite{RC} (also see \cite{witten}) which
arises in Riemannian geometry and is used in General Relativity \cite{wald,toolkit,skssg,AIRE,OLAS}.  
There too, the central aim is to analyse the  
behaviour of a bunch of trajectories
(geodesics). In General Relativity, or any metric theory of gravity,
a curved spacetime represents a gravitational field. Freely falling (no other non-gravitational forces) trajectories 
of test particles (massive or massless) are the {\em extremal} curves
or {\em geodesics} in a curved spacetime. A family of such non-intersecting geodesics
defines a {\em geodesic congruence}. It is therefore natural to ask--what happens to   
an {\em initially converging}  geodesic congruence?
The answer leads us to the {\em focusing theorem}
which states the following: {\em  under specific conditions (known in technical jargon as the convergence condition and 
the absence of vorticity/rotation), 
the family of geodesics must end up meeting ({focusing}) within
a finite value of a parameter $\lambda$ (similar to `time' in mechanics, $\lambda$ labels
points on the geodesics)}. Thus, focusing leads to a
{\em breakdown} of the definition of a  congruence. 

\noindent The term  {\em focusing} is 
quite commonly known and used in geometrical optics. Its usage here, is in the sense that trajectories
of mechanical systems may meet within a finite value of time. The meeting point is 
the {\em focus} or a {\em focal point/curve}. In optics, trajectories are
light rays and focusing may be related to the occurrence of caustic
curves where light intensities are enhanced drastically.
\cite{sachs, perlick}.

\noindent What does the focusing theorem signify in the
context of General Relativity? Since the Einstein field equations
relate geometry to matter, the {\em geometric} condition for focusing may be translated to that for
matter \cite{wald, toolkit}. Such a condition, simply stated,
is just the physical requirement of positive energy density.
Therefore, the attractive nature of gravity leads to focusing--
an almost obvious conclusion!
Further, it is possible that the focal point of a family/congruence 
is a spacetime singularity  
(eg. the big-bang or a black hole singularity, where one encounters
extreme spacetime curvature or infinite matter density).
Hence the role 
of the {\em Raychaudhuri equation and the focusing theorem} arise as
crucial ingredients in the
proofs of the celebrated singularity theorems 
of Penrose \cite{penrose} and Hawking \cite{hwp}, \cite{hawking}.

\noindent However, it is important to realise that focusing 
as such, can be completely benign.  
The focal point need not be a spacetime singularity, but only a point/curve
where the converging family of trajectories meet. This brings us back to the question asked in the first paragraph above--when does meeting happen, what are the conditions? It is this point of view (i.e. benign focusing) 
which we take forward while developing our analog  Raychaudhuri equation in
mechanics \cite{KTCM}.

\noindent In elementary mechanics, given a force field or a potential we can, from Newton's 
second law,
obtain a precise trajectory, once appropriate initial conditions are
provided \cite{kk,symon}.  To develop the approach highlighted above, we need to properly
define a family of trajectories as well as variables associated with
the collection, as a whole.
Once such variables 
are defined, their values at each instance of time
will show the nature of evolution of  
the family, as a single entity.  
The question of convergence/divergence or the meeting/avoidance of trajectories
in  finite time can therefore be addressed and linked to the
behaviour of such variables \cite{KTCM}.
Varying the initial positions and momenta (velocities) around 
specific values yields  different trajectories in configuration space.
One useful variable for a family is the gradient of the
velocity, known as the `expansion', which, as we will see, 
appears in the Raychaudhuri equation and is central to our forthcoming discussion.

\noindent Our article is arranged as follows. In the next section (Section II), we briefly outline, recalling earlier work \cite{KTCM}, the
theoretical model related to the behaviour of a family of trajectories in mechanics. The definitions of the meeting time
as well as the expansion are both introduced here. 
Thereafter, in Section III, we move on to the experiment, elaborating on various details of 
experimentation. We report on how the theoretical model tallies with our experimental findings in Section IV. A briefing on the 
correspondence between mechanical systems and relativity {\em vis-a-vis} the Raychaudhuri equation
and focusing, appears in Section V.
Finally, we conclude with remarks on possible future investigations.

\section{Theoretical model}

\noindent The equation of motion of a particle of unit mass, in the force field $f$ (with a potential $V$), in one dimension, is given as \cite{symon, kk}:
\begin{equation}
\dot{u}=\ddot{x} =-\frac{\partial V}{\partial x}.
\label{eq1.3}
\end{equation}
where, a dot and a double-dot denote, respectively, a first derivative and a second derivative w.r.t. time.
\noindent {Given the potential or the force field,
as well as initial ($t=0$, say) conditions on position and velocity
($x=x_0$, $u=u_0$ at $t=0$), one can write down the expression for
$x(t)$ (assuming integrability). A different set of initial
conditions ($x=x_0+\Delta x_0$, $u=u_0+\Delta u_0$ at $t=0$) will
result in a {\em different trajectory.} Thus, fixing the ratio of the difference in initial velocity and the difference in initial position of the trajectories
$(\frac{\Delta u}{\Delta x}$ at $t=0$ or $\frac{\Delta u_0}{\Delta x_0})$, we can generate a family of
trajectories for the mechanical system (see Figure \ref{drawingtrajectories}). Each trajectory in the family has a different initial position and initial velocity, but, for the collection,
the ratio $\frac{\Delta u}{\Delta x}$ at $t=0$ is fixed.

\noindent One may further ask about the behaviour of 
$\frac{\Delta u}{\Delta x}$ for this family, at different future time
instances.
This leads us towards the analysis of the kinematics of the 
family of trajectories, as a whole. We define a new variable 
$\theta(t)$ (essentially, the ratio $\frac{\Delta u}{\Delta x}$ at
each time, with a fixed initial value and w.r.t. a single reference trajectory) for the family. $\theta (t)$ is
named the expansion \cite{KTCM} and is defined as:
\begin{equation}
\theta(t)=\frac{\partial u}{\partial x}.
\label{eq1.4}
\end{equation}
It is obvious that $\theta(t)$ at each $t$, is the gradient of the velocity of neighboring trajectories in the family w.r.t. a single reference trajectory \cite{KTCM}. More rigorously one should write $\theta (x(t))$, i.e. its
time-dependence is through $x(t)$. We will however continue using $\theta(t)$ 
below with the understanding that it is actually $\theta(x(t))$.
\begin{figure}[h!]
\epsfxsize=4in\epsffile{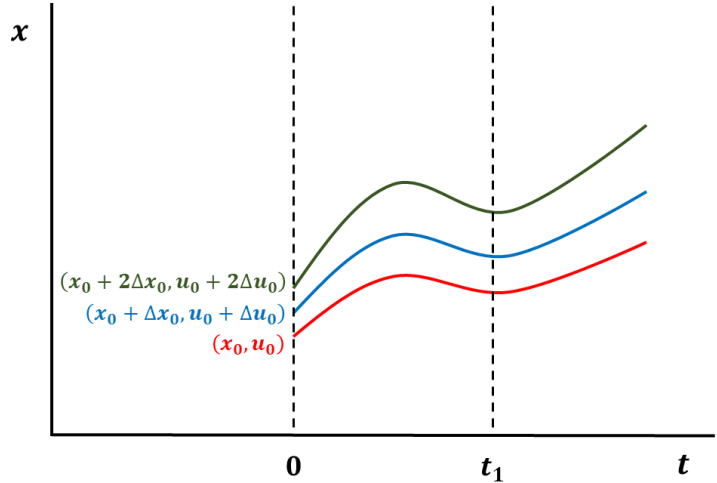} \caption{{ Qualitative plot of a family of trajectories (constructed w.r.t. the reference curve (red)) for an arbitrary system. Here, the initial position and initial velocity of the trajectories have been written in brackets with the same color at the left side of the $t=0$ dashed line as (initial position, initial velocity). The initial expansion (expansion $(\theta (t))$ at $t=0$) is $\frac{\Delta u_0}{\Delta x_0}$. At a later time (say $t_1$), the expansion ($\theta (t)$) is calculated by the ratio of difference in velocity and difference in position of the trajectories ($\frac{\Delta u}{\Delta x}$) at that time.}} \label{drawingtrajectories}

\end{figure}

\noindent Next, we ask, what is the differential equation obeyed by $\theta (t)$?
Since $\displaystyle \frac{d}{dt} \equiv \frac{dx}{dt} \frac{\partial}{\partial x}$ (see discussion later on using the convective derivative)
we have,
\begin{equation}
\frac{d\theta}{dt}=\ u\frac{\partial\theta}{\partial x} =\ \frac{\partial}{\partial x}\left(u\theta\right)-\theta^2, 
\end{equation}
and
\begin{equation}
u\theta=u\frac{\partial u}{\partial x} = \frac{du}{dt} = f_{ext} =-\frac{\partial V}{\partial x}, 
\end{equation}
where we have used the equation of motion (\ref{eq1.3}). Thus, the
final differential equation obeyed by $\theta(t)$ is,
\begin{equation}
\frac{d\theta}{dt}+\theta^2=\ -\frac{\partial^2V}{{\partial x}^2}.
\label{eq1.9}
\end{equation}
}
\noindent In some problems, we may not know a Lagrangian or the potential function
though the equation of motion may exist. This includes non-potential/non-conservative force fields \cite{KTCM}. In such cases, the equation for $\theta(t)$ is,
\begin{equation}
\frac{d\theta}{dt}+\theta^2=\ \frac{\partial f_{ext}}{\partial x}.
\label{eq1.10}
\end{equation}

\noindent One may question the usage of the directional derivative
$\frac{d}{dt} = u\frac{\partial}{\partial x}$, as opposed to the
convective derivative, i.e. $\frac{d}{dt} = \frac{\partial }{\partial t} + u\frac{\partial}{\partial x}$. It is easy to check that such a change does not
affect the final equation for $\theta$. In particular, assuming the Euler equation
in one space dimension,
\begin{equation}
\frac{\partial u}{\partial t} + u \frac{\partial u}{\partial x} = f
\end{equation}
instead of Newton's second law, one can obtain the same evolution equation
for $\theta$ (i.e. $\frac{\partial u}{\partial x}$). 

\noindent In the case of the pendula or simple harmonic oscillators (with
$f= -\alpha^2 x$), the Euler equation has a simple solution: 
\begin{equation}
u (x,t) =\alpha \, x \,
cot \left (\alpha t +\beta\right )
\end {equation}
which represents the velocity field ($\beta$ is a constant).
The integral curves of this velocity field (obtained from $\frac{dx}{dt} =u$) are,
\begin{equation}
x(t) = C \sin \left (\alpha t +\beta \right )
\end{equation}
where $C$ is a constant. One can relate the constants $\beta$ and $C$  to initial
conditions on $x$ and $u$, when choosing a
{\em specific, single curve} in the family. With a fixed initial $\theta$, one can
obtain its evolution as well as the evolution equation, directly
from the velocity field too. Alternatively, as in the preceding
discussion here, one may obtain the time evolution of $\theta$
(i.e. $\frac{\partial u}{\partial x}$) 
by transporting
$\frac{\partial u}{\partial x}$ in time starting from a fixed initial value. 
Both approaches eventually yield the same final result, i.e. the
same equation for $\theta$. Thus, it is clear that the time-dependence of $\theta$ is {\em through}
the time-dependence of $x(t)$, and is not explicit.

\subsection{Notion of meeting of trajectories}\label{subsec1.1}

\noindent {Let us now develop the notion of
meeting of trajectories.  As stated just above, the trajectory $x(t)$ has an initial position $x_0$ and initial velocity $u_0$. Further, name the trajectory with initial position $x_0+\Delta x_0$ and initial velocity $u_0+\Delta u_0$ as $x'(t)$.} One can write,
\begin{equation}
    \Delta u_0=\frac{\partial u}{\partial x}\bigg|_{(x=x_0, t=0)} \cdot \Delta x_0=\theta_0 \Delta x_0, \notag
\end{equation}
where $\theta_0 $ is termed as the initial expansion \cite{KTCM} and is given as:
\begin{equation}
    \theta_0= \frac{\Delta u}{\Delta x}\bigg|_{(t=0)}.
\label{theta0def}
\end{equation}
If at time $t=t_f$, the two trajectories meet, then,
\begin{equation}
x\left(t_f\right)=x^\prime\left(t_f\right). \notag
\end{equation}
From this condition, we can find the initial value of $\theta_0$ for which two trajectories may meet at some future time $t_f$. 
Notice that $\theta_0$ depends on the ratio of $\Delta u$ and $\Delta x$
at $t=0$. Thus, there
are infinite  possible values of $\Delta u$ and $\Delta x$ at $t=0$, which 
have the same $\theta_0$, thereby defining a family.
It may also happen that the family of trajectories never meet. 
In such a case, there is no finite value for the meeting time \cite{KTCM}.

\noindent In one dimension, we can also write $\theta\left(t\right)$ as,
\begin{equation}
    \theta\left(t\right)=\ \frac{\partial u}{\partial x}\approx\  \frac{1}{\Delta x}\ \frac{d\Delta x}{dt}.
\end{equation}
Thus, it may be interpreted as the fractional rate of change of separation between two trajectories \cite{KTCM}.
Further rewriting and integrating gives,
\begin{equation}
    \int\frac{\mathrm{d\Delta}x}{\mathrm{\Delta}x}=\ \ \int\theta\left(t\right)dt, \notag
\end{equation}
which implies,
\begin{equation}
{\Delta x=\Delta x_0\ .}{\mathrm{e}^{\int_{\mathrm{0}}^{\mathrm{t}}\mathrm{\theta}\left(\mathrm{t}\right)\mathrm{dt}\ }.}^{\mathrm{\ \ }}\mathrm{\mathrm{\ }}
\end{equation}
Hence, the divergence or convergence of the family of trajectories is related 
to the values of $\theta\left(t\right)$. If $\theta\left(t\right)\rightarrow-\infty$ in finite time, then $\Delta x\rightarrow 0$ in finite time, i.e., trajectories converge. This is
the well-known notion of meeting/focusing of a family of trajectories 
(see the penultimate
section where the analogy with geodesic congruences is discussed). 
If $\theta\left(t\right)\rightarrow \infty$ in finite time, then 
$\Delta x\rightarrow\infty$ in finite time, i.e. the family of 
trajectories diverge. Such behaviour is termed as defocusing.

\noindent We now move on to discuss the simple harmonic oscillator (or,
equivalently, simple pendula) \cite{KTCM}.
The equation of motion here is:
\begin{equation}
      \ddot{x}=-\ \alpha^2\ x,
\end{equation}
where $\alpha $ is the angular frequency.
The general solution turns out to be:
\begin{equation}
    x(t)=\ x_0\cos(\alpha t)+\left(\frac{u_0}{\alpha}\right)\sin(\alpha t),
\label{sho123}
\end{equation}
and the velocity $\dot x=u$ is given as,
\begin{equation}
    u(t)=\ {-\alpha x}_0\sin(\alpha t)+u_0\cos(\alpha t).
\end{equation}
The initial conditions  are $\displaystyle x(t)|_{(t=0)}=x_0,\ u(t)|_{(t=0)}=u_0$.

\noindent Let us assume we have two pendula with initial positions $x_0$ and
$x_0 +\Delta x_0$ and initial velocities $u_0$ and $u_0 +\Delta u_0$, 
respectively. Here $x(t)$ may be taken as the variable `length of the pendulum $\times$ angle'.
Using $x(t)$ (Equation \ref{sho123}), we obtain,
\begin{eqnarray}
x_1(t)=\ x_0\cos(\alpha t)+\left(\frac{u_0}{\alpha}\right)\sin(\alpha t), \notag\\
x_2(t)=\ (x_0+{\Delta x}_0)\cos(\alpha t)+\left(\frac{u_0+{\Delta u}_0}{\alpha}\right)\sin(\alpha t). \notag
\end{eqnarray}
Thus, the separation between neighbouring trajectories at time $t$ is,
\begin{equation}
\Delta x(t)=\Delta x_0\cos(\alpha t)+\frac{\Delta u_0 }{\alpha }\sin(\alpha t)=
\Delta x_0\left\{\cos(\alpha t)+\frac{\theta_0}{\alpha}\sin(\alpha t)\right\}.
\label{deltaxt16}
\end{equation}

\begin{figure}[h!]
\epsfxsize=4in\epsffile{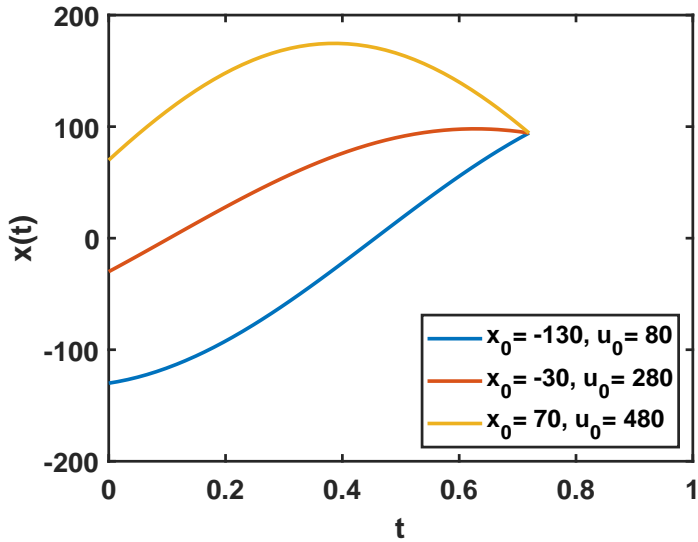} \caption{
Simple harmonic oscillator/simple pendula: Plots of trajectories  for arbitrarily chosen values of the initial position $(x_0)$ and the initial velocity $(u_0)$. {Note that $[\theta_{0}]$= T$^{-1}$ and $[\alpha]$=T$^{-1}$. Here, $\theta_0$=2, and $\alpha$ =3 in corresponding units.}}
\label{tsho1.1}
\end{figure}

\noindent Let the trajectories meet at time $t=t_f$ (see Figure \ref{tsho1.1}). We have,
\begin{equation}
x_1\left(t_f\right)=x_2\left(t_f\right)\mbox{ or } \Delta x(t_f)=0, \notag
\end{equation}
which gives, from {Equation \ref{deltaxt16}}, 
\begin{equation}\label{eq1.25}
    \theta_0\tan(\alpha t_f)=-\alpha.
\end{equation}
Thus, we obtain a relation between focusing time $\left(t_f\right)$ and initial expansion $\left(\theta_0\right)$. We can surely consider more than two pendula
and adjust their initial positions and velocities in such a way that
the initial value of expansion $(\theta_0)$ is the same for the family.

\begin{figure}[h!]
\epsfxsize=4in\epsffile{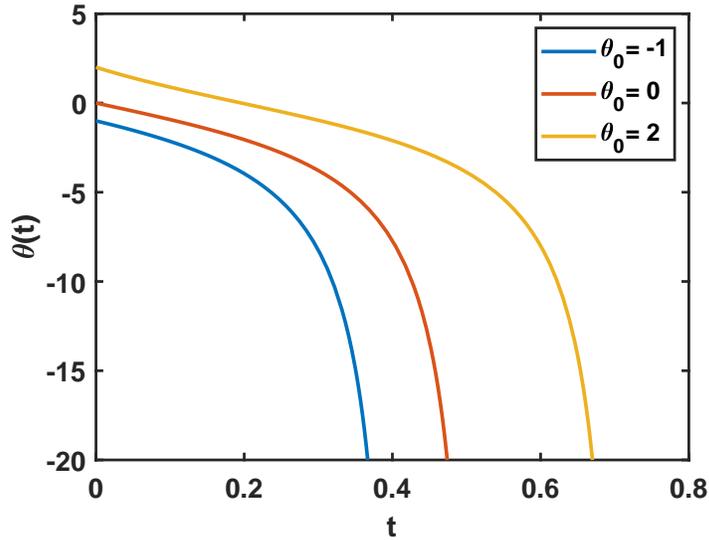} \caption{
Simple harmonic oscillator/simple pendula: Expansion $\theta(t)$ for various 
$\theta_0$ {in unit of T$^{-1}$ with $\alpha$=3 units}}
\label{tsho1.2}
\end{figure}

\noindent  Further, one may obtain the above expression for $t_f$ 
from the solution of the equation for $\theta(t)$ given as,
\begin{equation}
\frac{d\theta}{dt}\ +\theta^2=-\alpha^2.
\end{equation}
{It is straightforward to integrate this
simple first order differential equation.}
The solution turns out to be \cite{KTCM},
\begin{equation}\label{thetasho}
\theta\left(t\right)=\alpha\left(\frac{\theta_0-\alpha\tan(\alpha t)}{\alpha+\theta_0\tan(\alpha t)}\right),
\end{equation}
where the initial condition is $\displaystyle \theta(t)|_{(t=0)}=\theta_0$.
Figure \ref{tsho1.2} shows the variation of $\theta(t)$ with $t$.

\noindent Since the condition for meeting of trajectories in finite time is
$\theta\rightarrow-\infty$ as t$\rightarrow t_f$, we find,
\begin{equation}
\theta_0\tan(\alpha t_f)=-\alpha,
\end{equation}
which is the same as (\ref{eq1.25}).\\
Thus, the formula for the meeting time is given as,
\begin{equation}\label{sho234}
    t_f=\frac{1}{\alpha}\tan^{-1}{\left(\frac{-\alpha}{\theta_0}\right)}.
\end{equation}

\begin{figure}[h!]
\epsfxsize=4in\epsffile{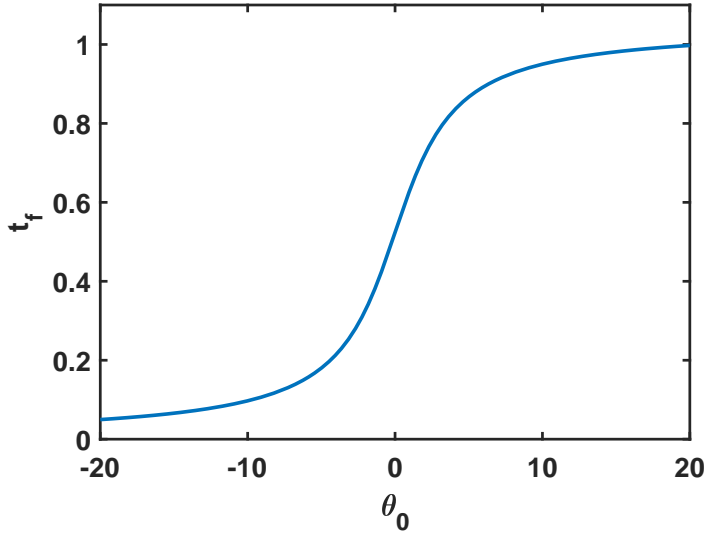} \caption{
Simple harmonic oscillator/simple pendula: Meeting/focusing time ($t_f$) with initial expansion{($\theta_0$ in T$^{-1}$ unit), $\alpha =3$ units.}}
\label{tsho1.3}
\end{figure}

\noindent As the tangent function can have any value between $-\infty$ to $\infty$, trajectories will meet for all values of the initial expansion ($\theta_0$) and $\alpha$. Figure \ref{tsho1.3} shows the variation of $t_f$ with $\theta_0$.

\section{Experiments}
\noindent We now set up an experiment involving simple pendula 
with the aim of learning whether our theoretical model and its quantitative predictions can
explain the experimental observations. In particular, the meeting/focusing time is one quantity which we obtain in our model and also measure in the experiment.

\subsection{Practical considerations}
\noindent A practical realization of the theoretical model, discussed above, demands a  modification in the
given expression for $t_f$ due to finite size of the bobs of the pendula and a  careful look at the issue of air-damping, which we elaborate below.
\subsubsection{Finite size-correction for meeting/focusing time}

\begin{figure}[h!]
\epsfxsize=3in\epsffile{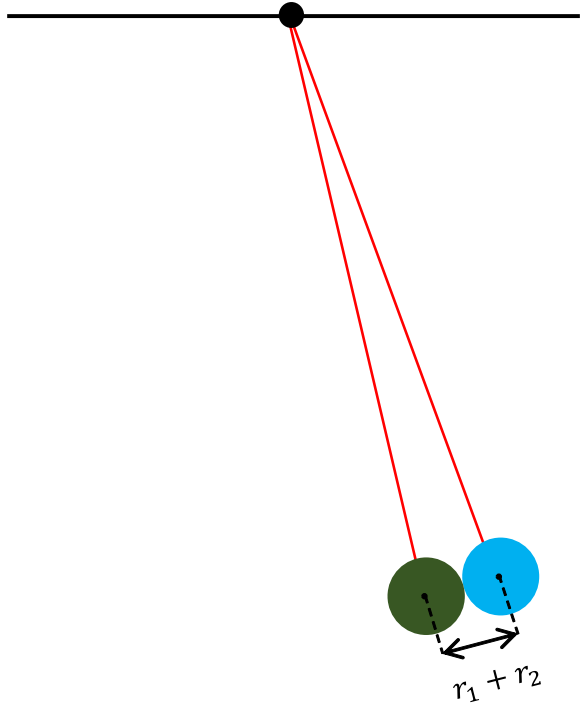} \caption{{Finite-size correction}}
\label{finite}
\end{figure}
\noindent  In Section II, the theory was developed assuming
point masses representing the bobs of the simple pendula. 
For practical purposes, the
bob of a simple pendulum is of finite size. In the derivation of
Equation \ref{sho234}, the focusing/meeting of two trajectories occur
when the separation between point particles becomes identically
zero. In practice, when two bobs meet, they do so at their
boundaries and not at  their centers of mass (see Figure \ref{finite}). 
Thus, Equation \ref{sho234} needs to be modified for this finite-size effect.

\noindent Let us assume that we have two bobs with initial positions
$x_{01}$, $x_{02}$ and initial velocities $u_{01}$, $u_{02}$
respectively. Following Equation \ref{sho123}, the positions of these
bobs after some time $t$ can be written as,
\begin{eqnarray}
x_1(t)=\ x_{01}\cos(\alpha t)+\left(\frac{u_{01}}{\alpha}\right)\sin(\alpha t), \notag  \\
x_2(t)=\ x_{02}\cos(\alpha t)+\left(\frac{u_{02}}{\alpha}\right)\sin(\alpha t),   \notag
\end{eqnarray}
where, $\displaystyle \alpha=\sqrt{\frac{g}{l}}$, and $l$ is length
of the pendula.
When they strike/meet, the separation between them
is equal to the sum of their radii, i.e.
\begin{equation}
{\big|x_2(t_f^\prime)-x_1(t_f^\prime)\big| = r_1+r_2},
\label{ex1}
\end{equation}
where, $r_1$ and $r_2$ are the radii of the bobs. Using the
above-stated
expressions for $x_{1}(t)$, $x_{2}(t)$ in Equation \ref{ex1}, we obtain
\begin{equation}
\big|(x_{02}-x_{01})\cos(\alpha t_f^\prime)+\left(\frac{u_{02}-u_{01}}{\alpha}\right)\sin(\alpha t_f^\prime)\big| = r_1+r_2
\label{finite1}
\end{equation}
Since the initial expansion $\displaystyle\theta_0=\frac{u_{02}-u_{01}}{x_{02}-x_{01}}$, we rewrite
Equation \ref{finite1} as
\begin{equation}
{\big|(x_{02}-x_{01})\big| \big|\cos(\alpha t_f^\prime)+\frac{\theta_0}{\alpha}\sin(\alpha t_f^\prime)\big| = r_1+r_2}
\label{finite_}
\end{equation}
Therefore, we have
\begin{equation}
\big| \cos(\alpha t_f^\prime)+\frac{\theta_0}{\alpha}\sin(\alpha t_f^\prime) \big |= A =
\frac{r_1+r_2}{\big|x_{02}-x_{01}\big|}
\label{finite0}
\end{equation}
where  $0<A<1$, which follows from the 
requirement $0<r_1+r_2 <\big|x_{02}-x_{01}\big|$ (i.e. initial separation
always greater than sum of radii).

\noindent Squaring both sides of Equation \ref{finite0} we obtain,
\begin{equation}
\left(\frac{\theta_0^2}{\alpha^2}+1\right)\sin^2(\alpha t_f^\prime)-\frac{2A\theta_0}{\alpha}\sin(\alpha t_f^\prime)+A^2-1=0
\label{finite4}
\end{equation}
which is a quadratic in $\sin(\alpha t_f^\prime)$ and 
can be easily solved to get $t_f^\prime$ as,
\begin{equation}\label{ex3}
t_f^\prime=\frac{1}{\alpha}\sin^{-1}{\left[\alpha.\left\{\frac{A.\theta_0
+ \sqrt{\left(\theta_0^2+\alpha^2-A^2.\alpha^2\right)}\ \ \ \ \
}{\theta_0^2+\alpha^2}\right\}\right]}
\end{equation}
The other solution of the quadratic, given as,
\begin{equation}
t_f^\prime=\frac{1}{\alpha}\sin^{-1}{\left[\alpha.\left\{\frac{A.\theta_0
- \sqrt{\left(\theta_0^2+\alpha^2-A^2.\alpha^2\right)}\ \ \ \ \
}{\theta_0^2+\alpha^2}\right\}\right]}
\end{equation}
is discarded since it gives a negative $t_f^\prime$ ($0<A<1)$.\\
\noindent Thus, in practice, the meeting/focusing time depends not only 
on the value of the initial expansion $(\theta_{0})$ and $\alpha$
(as in Equation \ref{sho234}), but also on the value of $A$.
In the limiting case when $r_1=r_2=0$ (or $A=0$), Equation \ref{ex3} reduces to Equation
\ref{sho234} (focusing time when the bobs are point particles).

\subsubsection{Air-damping}
\noindent {In reality, the oscillation of a pendulum is damped by air. To
calculate the damping constant $(\beta)$, 
we studied the successive oscillation of one of the two identical pendula, used in our main experiment. From the logarithmic decrement of the successive amplitudes (we have taken 20 oscillations) of the damped simple harmonic oscillator, the value of the damping constant $(\beta)$ is estimated to be  $(0.00309\pm0.00005)$  sec.$^{-1}$. This value is very small as compared to the angular frequency $(\alpha)$, which is $(3.260\pm0.002)$ rad. sec.$^{-1}$ (see next section for its measurement). Thus, in our experimental results, we ignored the contribution of air-damping while obtaining the
trajectories of the pendula.}

\subsection{Experimental Setup}\label{method1}

\begin{figure}[h!]
\epsfxsize=4.5in\epsffile{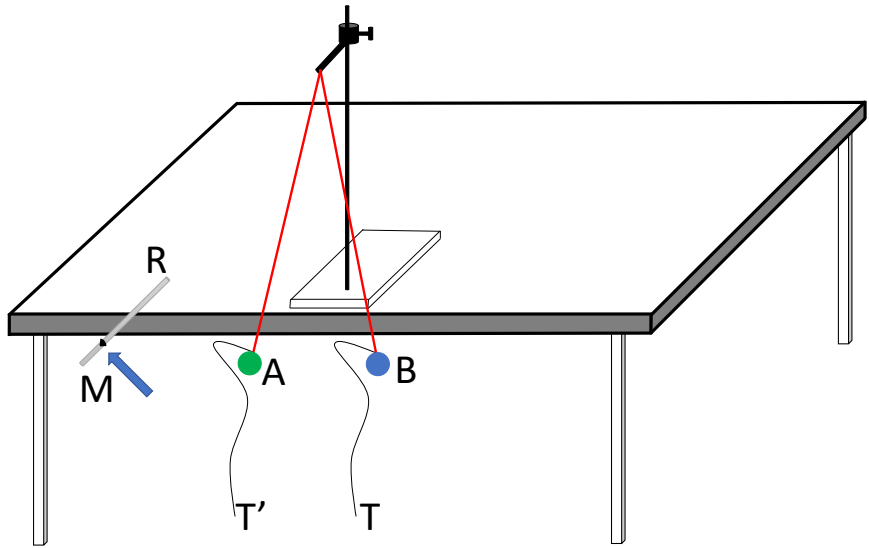} \caption{{Schematic
diagram of the experimental setup. A and B: Pendula, R:
Rod, M: marking on the rod and,  T and T': threads attached to the
bobs of the pendula.}} \label{setup1}
\end{figure}

{\noindent Figure \ref{setup1} shows the schematic diagram of our simple setup. In
our experiment, we took two pendula (A and B in Figure \ref{setup1}). The bobs of the pendula 
were of nearly equal
diameter (24.94$\pm$0.02 mm and 24.82$\pm$  0.02 mm) and weight (70.821 $\pm$ 0.001 
gm and
71.020 $\pm$ 0.001 gm). The length of each pendulum was nearly equal (92.2  $\pm$ 0.1 
cm).
{The angular frequency $\alpha$ can be found from the
length of each pendula and is, as stated above $(3.260\pm0.002)$ rad. sec.$^{-1}$.
We have also checked the angular
frequency from a direct measurement of the
time period of oscillation, and the difference between the value found
and that obtained from the length is
insignificant}.
To keep the motion of two bobs in the same plane, we fixed a
marked steel rod (R in Figure \ref{setup1}) on the table. The marking (M) on the rod and the equilibrium position of the bobs
were at the same distance from the side of the table. We fixed thin threads (T and T') on both bobs.
}

\subsection{Measurements}
\noindent To set the motion, threads T and T' were pulled at mark M and then released one after the other.
To achieve different initial velocities for two bobs, one of them was first released from the mark M on the rod. When it
attained a certain velocity in its trajectory, the bob of the other pendulum was released from the same mark M.
The time difference between the release of the first and second bobs was varied to
generate trajectories of the pendula with different sets 
of values for $x_{01}$, $x_{02}$, $u_{01}$, and $u_{02}$. {The amplitude of the oscillations 
in the experiment are small in order to conform to the
`simple' pendulum assumption.}
The trajectories of both bobs were recorded simultaneously using a wireless
camera that can record 60 frames per second.
It is to be noted that since we will be dealing with instantaneous positions and velocities of the two pendula, the origin of their trajectories is not relevant. 

\subsubsection{Measurement of position and time}
\noindent {To obtain the positions of the bobs, we processed the video using
OpenCV Python code \cite{py1, py2}. Two bobs were painted with
different colors, green and blue. After removing the
background using a bandpass filter, the program
differentiated two bobs following the HSV (hue, saturation and value) code.  
The code chose the planar projection of the spherical balls and defined
the center of the corresponding circles which determined the
position of their centers on their trajectories.
From the code, we obtained the position of the center of mass of the two bobs with respect to the equilibrium positions.
To convert it into a real physical unit of length, we measured the diameter of one of the bobs using slide-caliper and by the code.
The estimated
conversion factor, obtained from the ratio of these values, was used
throughout the experiment to study the trajectories of the bobs
in a real physical unit. This conversion factor also defines the
length corresponding to one pixel and is used as the error in position
measurements. }

\noindent As mentioned above, the video was recorded with 60 frames per second. By counting the number of frames between two desired positions of the bobs, the total time lapse could be estimated. The error in our measurement of time is 
$1/60$ second.

\subsubsection{Measurements of velocity}

\noindent 
We assume the motion only along  a line  (say x-axis), and get the
value of x-position at subsequent times using the code. Next, we use the
central difference
 method to determine its velocity (which is the derivative of position with respect to time).
 Note that, if we have the value of a function $f(x)$ at $x_i$, $x_i-h$ and $x_i+h$,
 then its first derivative at $x=x_i$ to first order of $h$, is given by,
\begin{equation}
f^\prime(x_i)=\frac{f(x_i-h)+f(x_i+h)}{2h}.
\end{equation}
In our case, with the known value of the position at time  $t_i$, $t_i-h$ and $t_i+h$, the velocity at $t=t_i$ to  first order in $h$, is given by,
\begin{equation}
v(t_i)={\dot x} (t_i)=\frac{x(t_i-h)+x(t_i+h)}{2h},
\end{equation}
where, a dot denotes a first derivative w.r.t. time. Note that we recorded the video with 60 frames per second. The positions can be obtained for each $h=\frac{1}{60}$ sec.

\section{Theoretical model versus experimental findings}
\begin{figure}[h!]
\epsfxsize=3.5in\epsffile{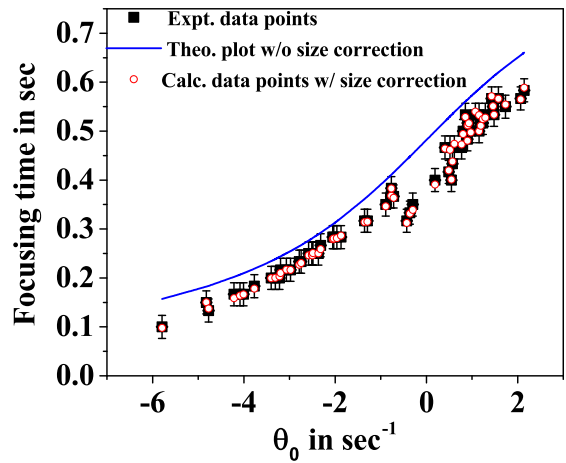} \caption{Variation of meeting/focusing
time with initial expansion.} \label{exp11}
\end{figure}
We now compare the
$t_{f}^{\prime}$ values found in our theoretical model 
and the experimentally observed meeting/focusing time for different
values of initial expansion ($\theta_0$). 
In Figure \ref{exp11} the
experimental data points are shown by square symbols. The solid blue
line plots the expected variation of the  focusing time without the
size correction following Equation \ref{sho234}. The red open symbols
represent values obtained from Equation \ref{ex3}, after including the
size correction to the meeting/focusing time. {It is to be noted that the
theoretical model based plot with the size correction is not a smooth curve as
the parameter $A$ in Equation \ref{ex3}  does not have a constant value.
It depends on the $x_{02}-x_{01}$ for a given value of
$\theta_0$ (Equation \ref{finite0}).} A fairly good agreement between the theoretical model based values and
the experimental data validates the predicted expression for $t_f^\prime$ (Equation \ref{ex3}).

\begin{figure}[h!]
\epsfxsize=4.5in\epsffile{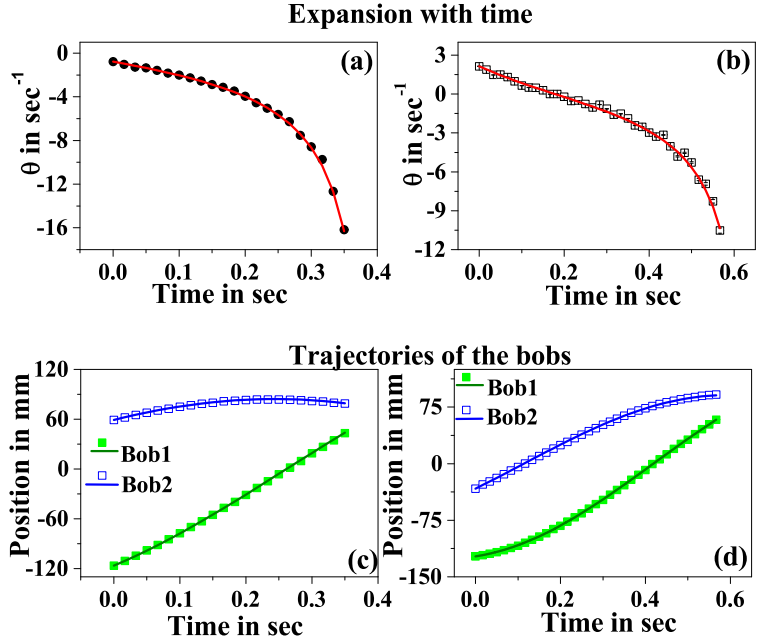} \caption{Variation of expansion
with time for (a) a negative value of $\theta_0$ (filled black
symbol), (b) a positive value of $\theta_0$ (open black symbol). The
calculated values of the same using Equation \ref{thetasho} are shown by solid  red lines. In (c) and (d), the
trajectories of two bobs corresponding to (a) and (b), respectively, are shown by green and blue symbols.  The
same as could be estimated from Equation \ref{sho123} are shown by the
solid lines.  The error bars for 
the experimental data are
within the size of the symbols.} \label{exp22}
\end{figure}

\noindent We also obtain the expansion of trajectories, $\theta(t)$, 
as a function of time, for both positive and negative initial values of $\theta$, 
i.e., $\theta_0$. The symbols
in Figure \ref{exp22} (a) and (b) plot the experimental data points
for a negative and  a positive value of $\theta_{0}$,
respectively. The theoretical model based plots following Equation \ref{thetasho}
are shown by the solid lines in these figures. The trajectories of the
two bobs for the above values of $\theta_0$ are shown in Figure \ref{exp22} (c) and
(d). Here too, the experimental data are shown by symbols in the
graphs, and
the theoretical model based trajectories following Equation \ref{sho123} are shown by the solid lines. It is important to note that the trajectories in
Figures \ref{exp22} (c) and \ref{exp22} (d) do not exactly meet. The reason for this is the
finite-size correction discussed in detail earlier. 
{In Figs. \ref{exp11} and \ref{exp22}, the theoretical curves are for $\alpha$=3.260 rad.sec$^{-1}$.  $x_0$ and $\theta_0$ are in mm and sec$^{-1}$, respectively.}

\section{Analogy}
\noindent As mentioned in Section I, the meeting/focusing time of trajectories derived  above
in the mechanics example and thereafter realised in an experiment, is analogous to the notion of 
focusing of geodesics in Riemannian geometry/gravitational physics. Let us now explain how the analogy works.
\cite{KTCM}.

\noindent To obtain the Raychaudhuri equations in Riemannian geometry,
we use the gradient of the normalised, timelike four-velocity field $u^i$  ($u_i u^i=-1$), given by $\nabla_j u_i$. Here $\nabla_j$ is the covariant derivative, $u_i=g_{ij}u^j$ and $g_{ij}$ is the component of the metric tensor in the line element $ds^2 = g_{ij} dx^i dx^j$.
$\nabla_j u_i$ is a tensor of rank two and can therefore be decomposed into its trace, symmetric traceless, and antisymmetric parts which represent, respectively,
the isotropic expansion, shear and rotation of the congruence (for
more details see \cite{wald,toolkit,skssg,AIRE}).
The trace of $\nabla_j u_i$, given as $\nabla_j u^j$ is defined as the expansion ($\theta$),
which, for one dimensional mechanics is just $\frac{\partial u}{\partial x}$ (Equation \ref{eq1.4}) \cite{AIRE}, a quantity we
have introduced and named as $\theta$. 

\begin{table}
\begin{tabular}{|c|c|}\hline
{\bf Mechanical systems} & {\bf General Relativity} \\ \hline 
Parameter: time (t) & Parameter: $\lambda$ (affine, non-affine) \\ \hline
Trajectories & Geodesics \\ \hline
Family of trajectories & Congruence of geodesics \\
& (timelike and null) \\ \hline
Meeting of trajectories & {Geodesic focusing} \\ \hline
Time of meeting & {Value of parameter $\lambda$ at focal point} \\ \hline
$\theta(t) =\frac{\partial u}{\partial x}$ & Expansion (Trace ($\nabla_j u_i$))\\
\hline
Equation for $\theta(t)$ (one dimension) & Equation for $\theta(\lambda)$ 
(three dimensions, timelike)\\
$\frac{d\theta}{dt} + \theta^2 = -\frac{\partial^2 V}{\partial x^2}= {k}(t)$ &
$\frac{d\theta}{d\lambda} +\frac{1}{3} \theta^2 =-\sigma^2 +\omega^2 - R_{ij} u^iu^j=g(\lambda)$
\\ \hline 
\end{tabular}
\caption{The analogy summarised.}  
\label{table1}
\end{table}
\noindent The Raychaudhuri equation for the expansion of a timelike geodesic congruence follows from an evaluation
of $u^k \nabla_k \left (\nabla_j u_i \right )$, which is the 
generalisation of the quantity $u\frac{\partial}{\partial x}
\left (\frac{\partial u}{\partial x} \right )$, for higher dimensions and curved spacetimes \cite{RC,wald,toolkit,skssg,AIRE}.
The equation for $\theta$ is given as:
\begin{equation}
\frac{d\theta}{d\lambda}+\frac{1}{3}\theta^2=-R_{ij}u^iu^j-\sigma_{ij}\sigma^{ij}+\omega_{ij}\omega^{ij}=g(\lambda)
\label{RE}
\end{equation}
where, $\lambda$ is a parameter, $\omega_{ij}$ is the antisymmetric
rotation tensor, $\sigma_{ij}$ is the symmetric traceless shear tensor and $R_{ij}$ is the Ricci tensor (for a definition of the Ricci tensor and a derivation of the
above equation see \cite{wald,toolkit}). 
\noindent The focusing theorem follows by assuming $\omega_{ij}=0$ and $R_{ij} u^i u^j \geq 0$, reducing the
equation to an inequality $\frac{d\theta}{d\lambda} +\frac{1}{3} \theta^2 \leq 0$. Integrating the inequality leads to the
conclusion that $\theta\rightarrow -\infty$ within a finite 
$\lambda$ \cite{wald, toolkit}.

\noindent Recall that the equation for the expansion ($\theta$) in one dimensional mechanical systems was (as shown in Equation \ref{eq1.9}),
\begin{equation}
\frac{d\theta}{dt}+\theta^2=\ -\frac{\partial^2V}{{\partial x}^2}=k(t)
\label{theta}
\end{equation}
Looking at Equation \ref{RE} and Equation \ref{theta} one can
easily notice similarities. In Equation \ref{RE}, $\lambda$ is a parameter and in Equation \ref{theta}, $t$ is an external parameter. In the second term of 
Equation \ref{RE} there is a $\frac{1}{3}$ factor arising due to three
space dimensions. In Equation \ref{theta} this factor is just one, 
as we work in only one space dimension. Finally, the R.H.S. term is a function of the parameter ($t$ or $\lambda$) 
in both equations. It may be noted that in mathematics, such equations are known as
Riccati equations \cite{riccati}.
Thus, the parallels between (a) the various quantities, (b) the equation for mechanical systems and the one in Riemannian geometry and (c) the resultant notion of meeting of trajectories/focusing are quite easily seen.  
We do seem to have
an {\em analog Raychaudhuri equation} in mechanics. 
The analogy is summarised in Table \ref{table1}.

\noindent How is such an analogy useful? 
First and foremost, it is a tool to introduce the
basic elements of the Raychaudhuri equation and the focusing
theorem to those who may not be familiar with it. 
Next, it is possible that the analog equation (as well
as its higher dimensional generalisations in mechanics),
can be investigated as an equation in its own right
with the motivation of
learning about collisions as well as avoidance in a family
of trajectories. Finally, solutions of the analog equation in mechanics
can lead us towards developing criteria on initial conditions
for invoking avoidance in a family of trajectories. Such 
an approach (though not carried out yet) may be relevant in the context of conjunction assessment and risk analysis 
programmes associated with artificial satellites, 
as explained in \cite{nasa}.  

\section{Concluding remarks}

\noindent Let us now conclude with some possible
avenues of future work.

\noindent It is certainly possible to go beyond the simple experiment discussed
here. A straightforward extension is to study systems where drag forces
are present. For example, the simple experiment related to Stokes' law \cite{kk} may be modified to perform a study similar to what has been done here. The theoretical 
model and its details appear in \cite{KTCM}. Further, one may
go beyond one dimension. Here too the theory has been developed
\cite{KTCM} and applied to projectile motion \cite{symon, kk}, which, may be
studied experimentally. The basic idea would be to shoot several
projectiles from different positions at different velocities
and obtain, using videography, the positions of each projectile
at subsequent times. One can study the evolution of expansion, shear and rotation in this example and find out how the meeting/focusing time varies with initial conditions.

\noindent {Moving away from mechanics, a formal study of families of trajectories also has useful applications in elasticity and fluids, as briefly indicated in \cite{toolkit}. More elaborate
discussion along these lines (especially the occurrence of caustics and vortices in media) may be possible following the detailed framework provided in \cite{cerveny1,cerveny2}.}

\noindent In conclusion, our present work is only a beginning. The 
future aim
is to broaden the scope of mechanics through studies on 
families of trajectories
as opposed to individual ones. The immediate outcome of these studies
is its direct link with a topic usually discussed in the context of
Riemannian geometry and General Relativity. It remains to be seen
whether such analyses have useful applications in mechanical systems.
At the very least, this novel approach surely provides a simple and 
worthwhile analog
which may be used while introducing the basics of the Raychaudhuri
equation and its consequences. Moreover, through the experiments 
reported here,
we have probably, for the first time, found a way to {\em realise in
a laboratory experiment} a rather involved concept like {\em focusing of trajectories},
through this analogy.

\section*{Acknowledgements}
\noindent We would like to thank Anang Kumar Singh and Kushal Lodha for their help in the experimental part. RPB thanks Department of Physics, IIT Kharagpur
where he was a student in the Master of Science programme, when this work
was carried out.

\end{document}